# Surface wake field model of beam-foil circular Rydberg states

Gaurav Sharma[1], N. K. Puri[1], Adya P. Mishra[2] and T. Nandi[3]

[1]*Delhi Technological University, Bawana road, Delhi 110042*
[2]*Atomic and Molecular Physics Division, Bhabha Atomic Research Centre, Trombay, Mumbai 400085*
[3]*Inter University Accelerator Centre, Aruna Asaf Ali Marg, New Delhi 110067*

Production of projectile Rydberg states in fast ion-solid collisions in H-like ions exhibits a pronounce target thickness dependence in spite of these states forming at the last layers. This occurs due to important role of the surface wake field which varies with the target foil thickness. Further, according to the proposed model Rydberg states with low angular momentum are transformed into a circular Rydberg states while passing through the field. The transfer occurs by a single multiphoton process with high probability depending upon the projectile ion velocity with respect to the Fermi velocity of the target electrons.

## I. INTRODUCTION

The electronic states of excited ions, atoms or molecules are divided broadly into two categories depending upon the extent of their orbits penetrating and non penetrating inside the core of electrons, which depends mainly on the angular momentum of the electron. Penetrating states are the ones for which the corresponding electron passes through the effective charge region of inner core electrons and the nucleus, while non-penetrating does not. The centrifugal barrier due to $l(l+1)/r^2$ does not allow electrons orbiting in high angular momentum to penetrate the core and therefore the electron is almost in an isolated state from the core, resulting in reduced perturbation, which increases thus the lifetime of the excited states [1]. Higher the value of $l$, lower the eccentricity of an orbit implies bigger size. The electrons in such high $l$ states are very less perturbed by the nuclear charge and therefore decay according to the selection rules for E1 transitions $\Delta l = \pm 1$. Core penetration for these states is small and the quantum defects are usually less than 0.1 [2]. In such conditions "$l$" is a good quantum number for describing the orbital electron properties. The non-penetrating states correspond to the maximum orbital quantum number ($l_{max}$) for a particular principal quantum number ($n$) have exactly a circular path of the electron and are thus called circular Rydberg states (CRS) [3]. Nuclear size correction is almost negligible for high $l$ states and zero for the CRS and thus improved Rydberg constant and energy levels can be determined very accurately. Apart from these, CRS can open various doors for many fields like determination of various relativistic and QED effects [4], different fundamental constants such as nuclear mass [5] and testing predictive power of a theory [6].

A lot of theoretical study and many experimental attempts were put forward for populating the electrons in CRS by various means. Freemen and Kleppner [7] proposed a Stark switching method to transfer electrons to high angular momentum states, then Koch and Martin [8] used laser excitation of collisionally excited Rydberg atoms, Hullet and Kleppner [9] proposed and used adiabatic rapid passages of time varying microwave fields, afterwards a RF-optical technique was suggested [10] but not realised yet. Delande and Gay [11] proposed a new theoretical scheme using weak magnetic field crossed to a weak time varying electric field, for exciting atoms to CRS with applicability to any atomic species which was later realised [12]. Liang et al. [13] and Cherlet et al. [14] used cross field method for levels up to $n=30$ while Breacha et al. [15] detected around $n=67$. In 1994 circularly polarised microwave field [16] on Na atoms was used to excite to CRS. All the above methods were applied to neutral atoms, where electrons were forced to CRS, while such states were never seen in any beam foil experiments worldwide, but recently Nandi [17] observed these states in H-like projectile ions and thereafter confirmed the same in projectile like ions [18]. Subsequently, Nedeljkovic et al. [19,20] explained these results using their two vector model. Further, Mishra et al. [21] also elucidated the above experimental results using the summing method of Curtis [22].

As the CRS can decay only through yrast chain due to the selection rules; they reach very late to the 2p state, the last candidate of the chain, from where Ly-α emission takes place in H-like heavy ions. Each delayed Ly-α transition originating from certain CRS appears as a hump at a particular delayed time, therefore it is possible to find the origin of CRS by summing the lifetime of the upper CRS levels in the decaying yrast chain. This method is applied recently in lifetime measurements [17] for H-like Fe and Ni projectile and various projectile like ions [18]. Lifetime measurements done in the past [23,24], has never detected the CRS. Examining the experimental methods used by Nandi [17,18] and others [23,24], we find that either the data have been taken for small flight times or the consecutive steps in the lifetime measurements are too large to notice the hump like structures [17,18], the characteristics of the CRS in the lifetime decay curves. As a result, decay curves following power law, the characteristics of the penetrating Rydberg states, are only detected, though both types of Rydberg states (circular and non-circular) populate in the beam-foil experiments. It may be worth noting here that H-like ions are the best candidate to study the CRS using beam-foil spectroscopy. Though He-like ions could be used to some extent, however, Li-like ions or higher multielectron systems do not allow to form CRS because of the autoionization process in such ions involving vacancy in the 1s shell. This is because of the fact that energetic ions are required to create highly charged ions and in this condition inner shell vacancy production is very high. This results in multiply excited states in multielectron systems containing $\geq 2$ electrons, which decay fast through autoionization before attaining to CRS [25].

Until this point we have discussed the formation of Rydberg states, now we take an attempt to examine the excitation mechanisms, which are responsible for such unique excitations. In order to explain the excitation mechanism two different thoughts are found in the literature. One supports that these states are formed inside the bulk of the foil as shown theoretically [26,27] and experimentally [28], while other suggests for the last layer effect from theoretical ideas [29–31] as well as experimental facts [32,33]. The theory of Day and Ebel [29] proposes that the wake-bound electron arriving in the vacuum, at its exit of foil, may be captured mostly into a hydrogenic bound circular orbits of the traversing ion. The experiment of Schiwietz et al. [33] using 140 MeV Ne beam on carbon foil finds very good agreement with classical-trajectory Monte Carlo calculations in the independent-particle model on quantum-state populations for ions interacting with gas as well as foil target indicating that last-layer electron capture is responsible for populating Rydberg states. Further, Mirkovic et al. [34], use two vector model (TVM) or surface phenomena model for explaining high $l$ population at the surface. The electron exchange occurs during intermediate stages of the ion-surface interactions; that results in the formation of the final Rydberg system. Nedeljković et al. [20] also apply the TVM for high $n$ and high $l$ to get the resonance like structures in agreement with experimental observations [17,18].

Though most of the findings supports the last layer effects, in contrast, a good experiment stated above [28] favours the bulk effect. By no means, such contradiction is acceptable; the experimental data can never be wrong, however, analysis probably can be faulty. Keeping this fact in mind, we have decided to reanalyse the data set for Ly-α ($2p\ ^2P_{1/2,3/2}$ - $1s\ ^2S_{1/2}$) and Ly-β ($3p\ ^2P_{1/2,3/2}$ - $1s\ ^2S_{1/2}$) x-ray cross sections reported as a function of foil thickness in Betz et al. [28]. Interestingly our analysis not only favours the surface layer effect for formation of the Rydberg states in beam foil excitations, but also it explores an unusual and important mechanism happening at the surface too. In this paper we plan to give a broad description on our findings.

**II. EXPERIMENT, ANALYSIS AND DISCUSSION**

In Betz et al. experiment [28], 125 MeV $S^{15+}$ or $S^{16+}$ ions were passed through carbon targets of thickness ranging from 2-200 μg/cm$^2$ kept at a certain distance from the detector to study the thickness dependence of the Rydberg state formation thoroughly. Ly-α and Ly-β x-rays of H-like S ions were recorded at $90^0$ to the beam axis using Si (Li) detectors with resolution 150 eV at 5.9 keV. Data were taken at four different distances 2, 4, 6 and 200 cm, so that only the long lived Rydberg states could be detected. Variations in the yields of Ly-α and Ly-β as a function of the foil thickness observed were attributed to the target thickness dependence of Rydberg states formed in the bulk of the foil. Let us reanalyse the data on different viewpoints.

The passage of ion beam through any solid medium leads to a distribution of charge states due to the various charge changing processes [35,36]. Hence, although $S^{15+}$ or $S^{16+}$ is being incident on the target foil many other charge states will be produced. However, since only Ly-α and Ly-β x-ray lines of H-like S are detected, the original states ought to be belonging to only $S^{15+}$. The generalised view of the theories [29–31], as well as experimental confirmation [32,33] stated above favours that the formation of Rydberg states happens right at the last layers. Therefore, high Rydberg states can be produced from $S^{15+}$ by various excitation processes near or at the surface; and from $S^{16+}$ by capturing electrons directly to such states at the surface. Hence, for either case, both $S^{15+}$ and $S^{16+}$ ionic states contribute to the formation of high Rydberg states in H-like ions irrespective of incident ion $S^{15+}$ or $S^{16+}$ and lower charge states ($q<15$) are not important. Very recently a thorough study of charge state fraction (CSF) distribution right at the bulk of the foil by Ly-α x-ray spectroscopy technique is carried out in our lab [37]. It shows that calculated CSF data from ETACHA code [38] represent the measured CSF distributions right at the collision region in the bulk of the foil very well in the energy range $\geq 2$ MeV/u. ETACHA code predicts that mostly CSF of incident charge state dominates in case of low foil thicknesses at 125

MeV energy, whereas CSF of $S^{16+}$ and $S^{15+}$ become comparable as the thickness increases.

Betz et al. [28] have measured both for Ly-α (2p $^2P_{1/2,3/2}$ → 1s $^2S_{1/2}$) & Ly-β (3p $^2P_{1/2,3/2}$ → 1s $^2S_{1/2}$) transitions with the $S^{16+}$ as incident ion and only Ly-α with the $S^{15+}$. Since, Ly-α transition is measured with both the incident ions; we have got an option to focus only on Ly-α measurements for the present analysis. Arguably the 2p $^2P_{1/2,3/2}$ state can be fed by both elliptic and circular states and feeding due to the elliptic states shows the power law decay [24], whereas the circular states display a hump like structure in the tail of the decay curves [17,18,39]. Since, Ly-α lines are detected at 20 mm distance or at t ~700 ps, the power law dependence of decay is highly ineffective [39]. Hence, Ly-α lines detected at such a long distance from the detector must be originating from the circular state ($n=7$, $l_{max}=6$) whose hump-like structure likely to appear around 700 ps. Thereby, we can discuss only the mechanism of CRS.

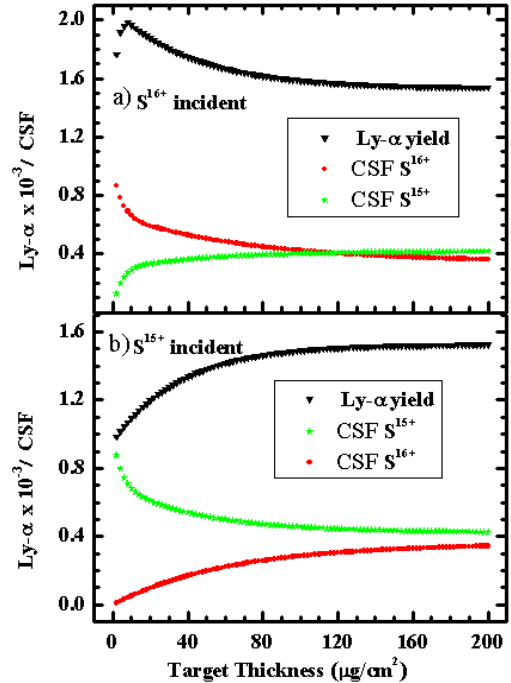

**FIG. 1.** Ly-α yield and CSF versus foil thickness plot. Absolute Ly-α intensities and CSF's of $S^{15+}$ and $S^{16+}$ as a function of the carbon foil thickness. The targets were kept at 2 cm away from the Si (Li) detector. CSF's are taken from the ETACHA code with the incident S ions (a) $S^{16+}$, (b) $S^{15+}$.

Ly-α yield for the incident ion $S^{16+}$ and $S^{15+}$ as a function of foil thickness are compared with CSF of $S^{15+}$ and $S^{16+}$ as obtained from the ETACHA code in Fig. 1(a) and (b). We note that Fig. 1(a) displays the scenario in case of $S^{16+}$ beam used as the incident ions, whereas Fig. 1(b) shows that for $S^{15+}$ used as the incident ions. In case of the lowest foil thickness, one can clearly notice that for the case of $S^{16+}$ incident ion CSF of 16+ is about 90% and that of 15+ is about 10%; in contrast for the case of $S^{15+}$ incident ions; CSF of 15+ is about 90% and that of 16+ is about 10%. Therefore, in the low foil thickness region $S^{16+}$ as outgoing ions play a major role while 16+ is used as the incident ions and $S^{15+}$ as outgoing ions is more important than $S^{16+}$ ions while $S^{15+}$ used as the incident ions. Figure 1(a) shows that Ly-α yield variation is fairly parallel to CSF of $S^{16+}$, which follows with the expectation in commensurate with the last layer mechanism. Looking on the Fig.1(a) more carefully one can notice that the parallelism holds good except low foil thickness region till 75 μg/cm$^2$. In case of incident $S^{15+}$, Ly-α yield variation is fairly parallel to CSF of the $S^{16+}$ with very large departure at the low foil thickness region (Fig. 1 (b)) as seen in the comparison between the Ly-α yield and the CSF of $S^{16+}$ in case of the incident $S^{16+}$ ions. This implies that even though formation of the Rydberg state is the last layer effect still the foil thickness plays a major role in both electron capture and excitation processes. The thickness of the last layers must be less than the mean free path of the collision system. The collisional mean free path can be calculated from $1/n\sigma$ where n=no of atoms per unit volume ~$10^{23}$ atoms/cc and σ = collisional cross section ~ $4\times10^{-18}$ cm$^2$, taking the formulations from Macquire et al. [40].

The size of the orbital, if it is formed in the bulk, must be less than the interplanar spacing of the amorphous carbon target. This distance is about 0.34 nm [41]. It implies that $n \leq 10$ orbitals are possible to form at the bulk of the amorphous carbon target. The range of $n$ formed at the surface layers can be $\geq 10$ and $l$ will mostly be $n$-1 for the present experimental conditions. However, all the orbitals $n > 10$ are not equally possible; some of the $n$ values only are possible [39].

### III. SURFACE WAKE FIELD MODEL

Laser excitation of Rydberg states is now a well established fact. The atom-laser interaction region which extends only a few Bohr radii around the atomic nucleus leads excitations to the Rydberg states [42]. Thereby, the electrons are excited to the Rydberg states by means of the electromagnetic waves. A recent study reveals that a finite electric field exists at the exit surface of the foil during the passage of ion beam [43]. One can obviously consider this field may play some role in producing the Rydberg states. Further, this study clearly hints on the thickness dependence of the surface wake field (SWF), which increases with the foil thickness and reaches to a saturation at a certain thickness. At this stage it is quite plausible to consider that variation of the SWF with the foil thickness may play an important role in forming the CRS in the last layers of the target foil. Such mechanism can be represented through a possible model as given below.

Before proposing the model we note that the energy loss in the bulk of the foil produces a strong electric field [44]. Vernhet *et al.* [45–47] have made thorough study on the Stark-mixing between $np$ and $ns$ states for $n$ up to 5 due to the wake field. The SWF induced Stark mixing between $np$ and $ns$ states is also shown to be important in our previous works [21,48] as mixing takes place between a short lived $p$ state and a long lived $s$ state. Nevertheless, the Stark mixing will make only a minor difference in case of CRS as both the participating levels have long lifetimes.

The surface energy loss exhibits that the positively charged ions suffer certain energy loss due to the SWF [43]. It means that the field attracts the ions towards the foil surface and repels the electrons away from the foil surface by an attractive field resulting from both, the bulk and the surface wake (Fig.1 of [49]). Let us make use of this fact here. Suppose the CRS are formed in the last layers in two steps: firstly, a high Rydberg state (elliptic) is formed due to electron capture or excitation and secondly, during the passage through the SWF the elliptic Rydberg state is promoted to the CRS because of the pull on the ions and push on the electrons. Figure 2 pictorially represents the proposed model for the excitation mechanism of the CRS. Stage I of Fig.2 represents an elliptic Rydberg state formed in the H-like ion at the last layers when the bulk wake field (BWF) is on and the SWF is off; electron is moving in the elliptical orbit $nlm$ ($l=|m|<n-1$) and the circular orbit remains vacant. The circular orbit shown is corresponding to a CRS $nl'm$ such that $l'=|m|=n-1$. As soon as the SWF is switched on adjacent to the foil surface the electron can be pushed to the CRS because of push-pull effect as shown in stage II and by this time ion reaches to the field free region, from this time onwards the electron remains in that orbit till its lifetime permits as shown in stage III. Range of the SWF is of the order of 5 nm [43], the time taken by the ion to traverse the field region is less than the time period of the orbiting electron. The strength of SWF is in the order of $10^7$ V/cm [48] and decreasing with distance from the foil [29]. Hullet and Kleppner [9] have created the elliptic Rydberg states by Laser-atom interaction in presence of the strong electric field and then passing these states in a time varying electric field. Hence, the similar experimental configuration is emulated in the ion-solid collisions in a conducting target. In this condition the interference effects favour all the states required from low $l$ to $l_{max}$ ($|m|=n-1$) can be simultaneously excited and superposition of all the states can be viewed as a single multiphoton event [9]. Here a worth noting point is that the probability of having the CRS from the elliptic Rydberg state depends on the magnitude of the SWF. As the SWF is small for low foil thickness, the probability of having the CRS from the elliptic Rydberg state is expected to be small. Thus, one observes the low yield of the Ly-α x-ray for low foil thickness as seen in Fig. 1.

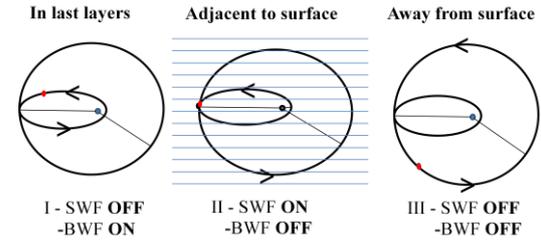

**FIG. 2.** Excitation model for CRS at the foil surface. I. NCRS is formed at the foil surface in absence of SWF and presence of BWF, II. NCRS is promoted to CRS in presence of SWF only, and III. CRS lasts as per its mean lifetime in the field free regions. Red dot indicates orbiting electron and black dot the nucleus of the ion. The electron can be at any point in the orbit while the ion exits the foil surface. The most favoured position is shown in the figure.

As can be seen in Fig. 1 that the CSF of 15+ and 16+ are comparable from 75 μg/cm$^2$ onwards upto 200 μg/cm$^2$. Whereas the contribution of 15+ and 16+ ions get reversed in case of S$^{15+}$ compared to S$^{16+}$ as incident ions as discussed above. Capture of an electron to S$^{16+}$ in high $n$ states at the surface layers leads to the Rydberg states in H-like S. Similarly excitation of the electron remaining with S$^{15+}$ to the Rydberg states may also be possible at the surface layers. Consequently, both the charge states i.e. sum of CSF of S$^{15+}$ and S$^{16+}$ may equally be responsible to produce the Rydberg states in H-like ions irrespective of the specific incident charge state used. However, the CSF varies with the foil thickness, we need to exclude the variation of charge state from x-ray photon yield data as a function of the target thickness. In order to do that we define a parameter R:

$$R = \frac{Ly\ \alpha\ yield}{CSF\ (S^{15+} + S^{16+})}$$

This quantity represents the relative CRS formation probability. Thus, the ratios so obtained with the incident ions S$^{16+}$ as well as S$^{15+}$ are plotted against the foil thickness as shown in Fig. 3(a) and (b), respectively. One may clearly notice that the Ly-α data show certain variation with the foil thickness at low thickness regions similar to the picture shown in Fig. 1. This can be attributed to the effect of SWF as discussed above, hence revealing the model predictions. Both the curves, 3(a) and 3(b), reach saturation at certain foil thickness. However, the Fig. 3(a) displays an overshoot from the saturation at certain foil thickness, in contrast such departure is not seen in Fig. 3(b). At this juncture we fail to find any possible reason. Whatsoever, the curve in Fig. 3(a) attains the saturation ($1.92 \times 10^{-3}$) faster about 32 µg/cm$^2$ than that in Fig. 3(b) ($2 \times 10^{-3}$) about 55 µg/cm$^2$. It is known that contribution of self-wake [50] in the SWF [43] is important in addition to the collective plasmon excitations. The data given in Table 1 in [43] clearly indicate that the SWF varies with the foil thickness. Further, the self-wake depends on the charge state of the incident ions [50], thus the saturation is achieved with the lower foil thickness while $S^{16+}$ is incident compared to $S^{15+}$.

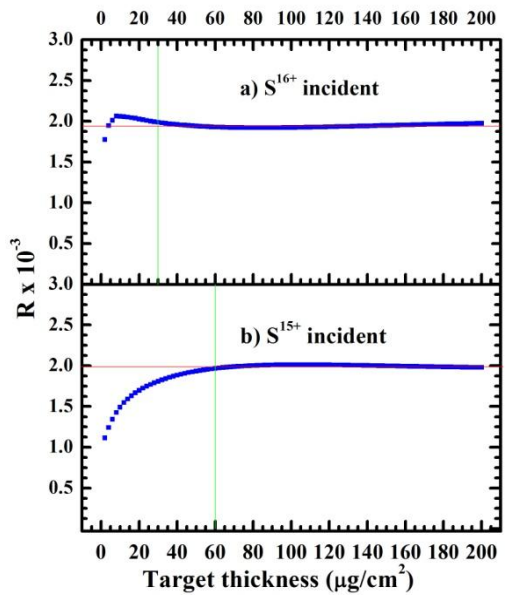

**FIG. 3.** The ratio R, as defined in the text, is plotted as a function of carbon foil thickness for incident ions (a) $S^{16+}$ and (b) $S^{15+}$. Red lines represent the saturated value of R and green lines represent the thickness where the saturation is reached.

### IV. CONCLUTION

To conclude, we have reanalyzed the Ly-α yield versus the foil thickness data [28] in the light of the facts that the ETACHA code represents the charge state distribution of the projectile ions in the bulk of the target foil [37] and the SWF varies with the foil thickness [43]. Though the Rydberg states in fast-ion foil collisions are formed at the last layers, the production of the projectile Rydberg states in H-like ions exhibits a pronounce target thickness dependence after having excluded the contribution from the charge variation with the foil thickness. Such a contradicting occurrence is explained clearly with the present model. The observed variation is due to the fact that the SWF varies with the foil thickness. Further, the Rydberg states with low |m| states are transformed to the CRS while passing through the SWF. The transfer occurs in the strong SWF as a single multiphoton event with a high probability depending upon the ion velocity with respect to the Fermi velocity of the target electrons.

The study of the ion–solid collisions provides an important connection between atomic physics and condensed matter physics. Many a time if some observed facts in the ion-solid collisions are not explicable with the atomic processes then that is attributed to the solid state effects, for example [45]. In this work, we have pinpointed that the solid state effects playing an intriguing role on populating the CRS is in fact the SWF.

### ACKNOWLEDGEMENTS

G. Sharma acknowledges University Grant Commission for financial and IUAC for all other supports.